# Temperature-tuning of near-infrared monodisperse quantum dot solids at 1.5-μm for controllable Förster energy transfer

*Ranojoy Bose[1], James F. McMillan[1], Jie Gao[1], Kelly M. Rickey[1], Charlton J. Chen[1], Dmitri V. Talapin[2], Christopher B. Murray[3], and Chee Wei Wong[1]\**

[1]Optical Nanostructures Laboratory, Center for Integrated Science and Engineering, Solid-State Science and Engineering, and Mechanical Engineering, Columbia University, New York, NY 10027, [2]Department of Chemistry, University of Chicago, Chicago, IL 60637, [3]Department of Chemistry, University of Pennsylvania, Pennsylvania, PA 19104

*rb2261@columbia.edu

**Abstract:** We present the first time-resolved cryogenic observations of Förster energy transfer in large, monodisperse lead sulphide quantum dots with ground state transitions near 1.5 μm (0.83 eV), in environments from 160 K to room temperature. The observed temperature-dependent dipole-dipole transfer rate occurs in the range of $(30-50 \text{ ns})^{-1}$, measured with our confocal single-photon counting setup at 1.5-μm wavelengths. By temperature-tuning the dots, 94% efficiency of resonant energy transfer can be achieved for donor dots. The resonant transfer rates match well with proposed theoretical models.

Lead chalcogenide (PbS and PbSe) quantum dots (QDs) have emerged as exciting candidates for quantum optics, in particular due to variable synthesis that allows for precisely tunable emission at the important optical communication wavelengths (1.3-1.55 μm) [1-3]. These QDs are also remarkable materials for solar photovoltaics, capturing the infrared spectrum [4]. Dispersion of dots in different solvents allows for their integration into infrared optical elements in a controllable post-fabrication process and is especially useful for several applications. For example, the QDs can be dispersed in PMMA and integrated into thin uniform films using electron beam lithography on silicon CMOS devices for cavity quantum electrodynamical (cQED) experiments [5, 6], and coated on microcapillary whispering gallery mode resonators for lasing at 1.53 μm [7].

The possibility of controllable quantum dot assembly also provides opportunities to explore interdot interactions through energy transfer mechanisms. Incoherent Förster resonance energy transfer (FRET) [8-11] can occur between different-sized quantum dots in monodisperse or mixed QD assemblies when dots are in close proximity to each other and there is sufficient overlap between the donor emission and acceptor absorption spectra. Resonant energy transfer is important in a wide-variety of applications, especially in the near-infrared region where applications include solar energy conversion as well as quantum-communication. In quantum dot systems, the dot-sizes pose a fundamental limit on the rate of Förster energy transfer. FRET has been observed in colloidal cadmium sellenide (CdSe) [12-15], cadmium sulphide (CdS) [16], indium phosphide (InP) [17] and, PbS QDs [18] at wavelengths below 1100 nm and often at room temperature. In all these cases, small dot sizes allow for assemblies where dots are in close proximity (<6 nm) to each other and fast Förster processes can occur. At 1.5 µm wavelengths, the dots are larger (5.5 nm diameter) resulting in a larger interdot distance, and slower Förster processes. Experiments are further complicated with the availability of fast sensitive detectors beyond silicon absorption wavelengths (longer than 1100 nm). In recent experiments, Förster energy transfer in quantum dot solids is demonstrated through a direct measurement of photoluminescence lifetime for small and large quantum dots. In this work we examine incoherent dipole-dipole interactions in large infrared quantum dot assemblies emitting in the near-infrared communication wavelength regions (1.3-1.7 µm), and demonstrate controllable Förster energy transfer rates by temperature-tuning our sample between 160 K and room-temperature. Our measurements match well with theoretical predictions, and the measured resonant transfer times range between 148 ns at 295 K to 30 ns at 160 K for dots emitting at 1450 nm in our 5.5-nm nanocrystals with <10% dispersities.

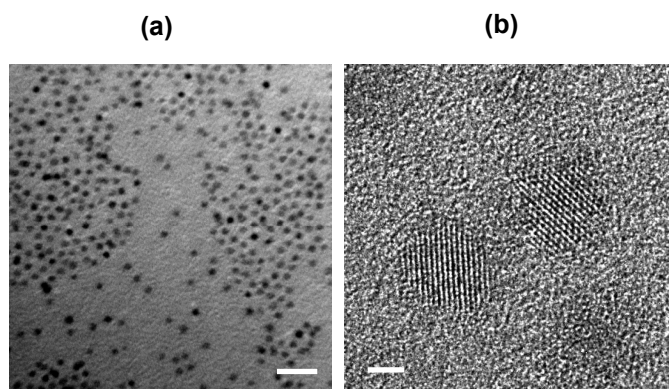

Figure 1 (a) Transmission electronic micrograph (TEM) of PbS QDs shows excellent size dispersity. These QDs have photoluminescence peak at 1475 nm. (b) High-resolution TEM of two PbS quantum dots in close proximity. Scale bar represents 50 nm in (a) and 2 nm in (b).

The PbS QDs in our studies (Figure 1) are synthesized by methods described elsewhere [1, 2] and are initially suspended in chloroform solution. The sample used exhibits room-temperature ensemble photoluminescence spectra centered at 1460 nm (solution) and 1475 nm (dried film; Figure 3a) and has an estimated quantum yield of 20% in solution. A small amount of the dense solution is dropped on a silicon wafer and allowed to dry, thereby producing a thick QD solid for Förster transfer studies. The observed red-shift in QD emission in the dried solid can be directly attributed to Förster energy transfer processes [12, 18]. Excellent photostability of our 5.5 nm PbS quantum dots [19] enables us to do experiments in an ambient environment at room temperature, and the dots remain remarkably stable over the temperature cycling in the cryostat.

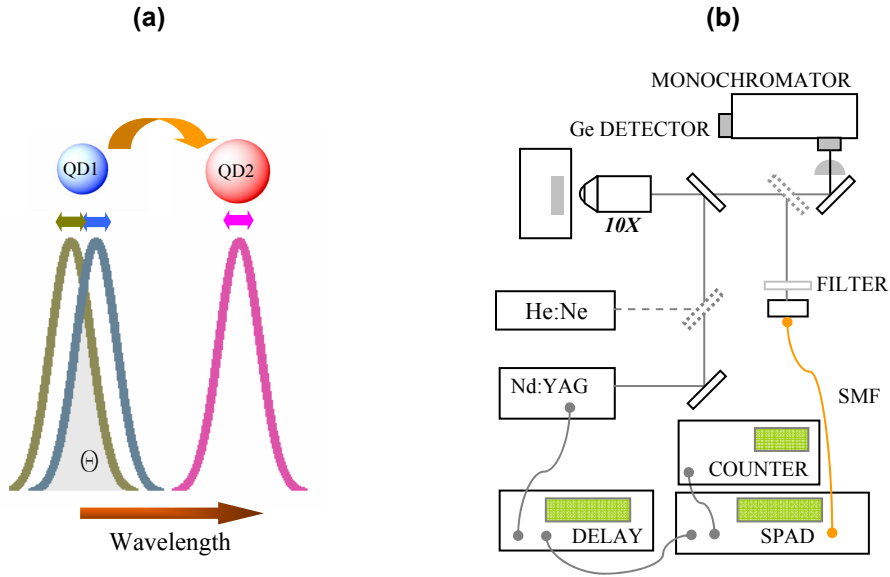

Figure 2 (a) Schematic of the Förster energy transfer process in the case of two quantum dots: The smaller quantum dot (high energy) may non-radiatively transfer energy to the larger quantum dot (lower energy) when there is sufficient overlap between donor emission (green curve) and acceptor absorption (blue), shown as the gray area under the curve. The acceptor emission is shown in purple. Both the resonance and linewidths of quantum dots change with temperature. (b) Schematic of the experimental setup: For PL measurements, a continuous wave He-Ne laser is used to pump the quantum-dot solid in a cryostat, and the collected emission is dispersed by the monochromator and collected at a liquid nitrogen-cooled Ge detector. For the lifetime measurements a pulsed Nd:YAG laser (2 ns, 15 kHz) is used to pump the solid, and also to trigger the single photon detector. The QD emission is collected in a single mode fiber, detected by the single photon detector and registered at the photon counter.

We first characterize the temperature-dependent photoluminescence of our PbS quantum dots. In the cryogenic measurements of the PL spectra, a 30 mW HeNe laser is used to pump the film of closely-packed PbS QDs in a Janis liquid-helium cryostat, with a 2.5 um spot size using a 10× objective lens with a 0.25 numerical aperture. The QD emission is collected using the same lens, dispersed with a 32 cm JY Horiba monochromator (Triax 320), and detected at a cooled Ge detector (Figure 2b). The cryostat temperature is varied between 160 K and 295 K. At room temperature, the PL spectrum (Figure 3a) extends from 1300 to 1700 nm, with a peak at 1475 nm (Stokes-shifted by ~ 100 nm) and full-width-at-half-maximum (FWHM) of 155 nm. At 160 K, the observed FWHM is 150 nm, and the peak is red-shifted to 1490 nm, (dE/dT=60 µeV/K). The observed shift of PL spectrum toward lower energy with decreasing temperature match well with observations for PbSe dots of smaller size [20, 21], as well as thiol-capped PbS quantum dots in glass [22]. In solution, the measured PL spectrum is affected by solvent absorption, resulting in unusual features. Further studies are required for an accurate explanation of the observed asymmetry in the QD solid.

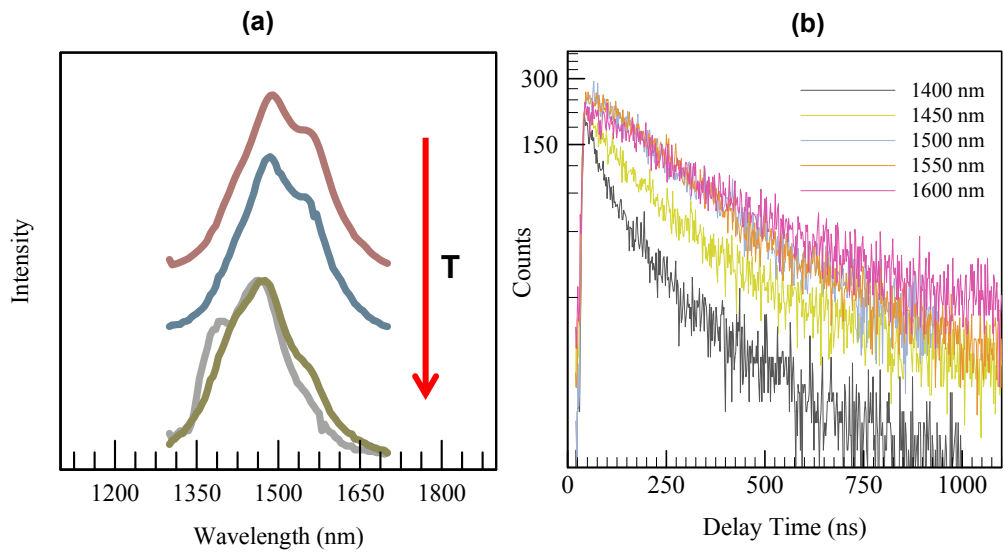

Figure 3 (a) Measured PbS QD photoluminescence spectrum in solution (gray) and solid at 160 K (red), 200 K (blue), and 295 K (green) (blueshifted with increasing temperature, and normalized at the peaks), characterized with a 30 mW average-power HeNe laser. The average change in the PL peak over this temperature range is measured to be 111 pm/K. (b) PL lifetime spectra for QDs in solid at various wavelengths on the PL curve (room temperature, 800 µJ/cm² excitation, corresponding to $\langle N_{e-h} \rangle < 1$ exciton/dot) showing evidence of Förster energy transfer, through a fast quenching of small dots and increased emission for large dots immediately after excitation. The measured Förster transfer rate at 1400 nm is $(60 \text{ ns})^{-1}$. All spectra are normalized for clarity.

For the time-resolved measurements of the photoluminescence lifetime, the pump is a passive Q-switch 2-ns 1.064 µm Nd:YAG laser with 12 uJ pulse energy at 15 kHz repetition rates. The 15 kHz repetition rate ensures complete relaxation of the samples between pulses. The pump has a synchronization photodiode to reduce jitter between the optical and electrical pulses to ~ 10 ps, and is attenuated to a fluence of 800 µJ/cm2 incident on the QD samples. The collected emission is coupled into a single-mode fiber, and detected by a Geiger-mode TE-cooled InGaAs/InP single photon avalanche diode (SPAD) from Princeton Lightwave with the necessary control electronics, bias and afterpulse suppression for 25% detection quantum efficiency and $4\times10^{-5}$ dark count probability for a 1.28-ns gate width. The SPAD gate width is set at either 5.12- or 10.24 ns, delay-triggered with the pump excitation laser, and the detection integrated over 0.5 to 4 s, depending on the detected counts. The total timing jitter in the system is estimated to be sub-nanosecond. The losses in the collection path are estimated to be about 55 dB, consisting of 10 dB from the QD to the first objective lens, 10 dB from the first objective lens – to – fiber collimation, 3 dB from the fiber-to-SPAD, 10% QE at detector and 10 to 20 dB due to the gating process. For the quantum dots used in this study, a monoexponential lifetime of 757 ± 10 ns at room temperature is observed with a dynamic range of 30 dB (Figure 3a), slightly shorter than the ~ 1 to 2 µs lifetimes observed by other groups [18, 23, 24] but longer than ~ 250 ns theoretical estimates based on the simple picture of dielectric screening [24]. In the solid film, such as shown in Fig. 2a, QDs are present in close proximity to each other, and the PL lifetimes are modified due to inter-dot interactions [25] and possibly reduced quantum efficiency ($\eta = \tau_{PL}/\tau_R$, where $\tau_{PL}$ and $\tau_R$ are the photoluminescence and radiative lifetimes of the quantum dots respectively). The measured lifetime in the solid ranges between 310 ns at room-temperature to 520 ns at 160 K.

Time-resolved measurements of the ensemble monodisperse PbS QDs in the solid at room temperature show a ~310 ns exponential decay with a dynamic range of 30 dB. The PL lifetime is observed to be a property of the solid, and may be considerably different for different dried films derived from the same solution. In order to study possible Förster energy transfer at room temperature, we use 12 nm linewidth pass filters at 50 nm increments from 1400- to 1600 nm (3-dB transmission) to probe spectrally-resolved lifetimes of dots of different sizes (Figure 3b). When focused, the laser pump fluence results in an occupation of less than 1 exciton per dot (maximum $<N_{e-h}> \approx 0.1$ excitons/dot). In the solid, a considerable dependence of the lifetime spectra on the energy emission is observed compared to dots in solution, where no such dependence is observed. For dots emitting at 1400 nm, a fast Förster energy transfer quenching is observed for the first 80-100 ns (decay constant ~ 60 ns), and a slower intrinsic decay is observed thereafter. The fast lifetime component becomes longer with a decrease in the exciton energy, and eventually approaches the ensemble lifetime ($\Gamma = \Gamma_{PL} + P(\lambda)\Gamma_{et}$ [12] where $P(\lambda) = \int_{\lambda}^{\infty} \Phi(\lambda')d\lambda'$ represents the probability of a viable acceptor and is derived from the spectral lineshape $\Phi(\lambda)$ for dots in solution, and $\Gamma_{et}$ is the measured transfer rate). From the measured lifetimes, we conclude that efficient energy transfer occurs when the acceptor bandgap is larger than the donor by at least $\Delta\lambda_{FRET}$ =100 nm at room temperature ($P_{REAL}(\lambda) = \int_{\lambda+\Delta\lambda_{FRET}}^{\infty} \Phi(\lambda')d\lambda'$), similar to the typical Stokes shift observed in these samples. However, the large inhomogeneous bandwidth of PbS QDs (~400 nm) allows for resonant energy transfer in monodisperse assemblies due to the availability of adequately-shifted acceptor dots within the ensemble spectrum. At 1500-, 1550-, and 1600 nm, the luminescence has a monoexponential decay constant of 250-, 308- and 310 ns respectively. At 1600 nm, an increase in photon counts is observed immediately after excitation due to the nonradiative transfer of excitons into the larger sized dots. This transfer occurs over a time of 50-100 ns at 1550-1600 nm. The observed results reflect the averaged Förster energy transfer over a large number of dots.

The observed transfer rate of (60 ns)$^{-1}$ is lower compared to observations in monodisperse CdSe (1.9 ns)$^{-1}$, and can be most reasonably explained through the expected increase in the interdot separation for PbS quantum dots, averaged over all donors at 1400 nm. As in the case of Ref. 12, the shorter lifetime observed from dots at the high energy end of the spectrum is attributed to the availability of a large number of large (low energy exciton) dots for Förster energy transfer (which results in emission at rates faster than 'normal' radiative recombination). The eventual slow decay is attributed to the decrease in the number of available dots due to carrier occupation, resulting in relaxation of the small dots at the ensemble lifetime.

The energy transfer rate between two distinct dipoles (in units of ps$^{-1}$) is given by $\Gamma_F \approx 1.18 V^2 \Theta$ [12, 14, 15, 26]. Here V is the coupling strength (in cm$^{-1}$ units) between donor and acceptor, given by $V \approx 5.04\mu_A\mu_D\kappa R^{-3}$, where R is the interdot distance, $\kappa$ is an orientation factor and $\mu_A$ and $\mu_D$ are the acceptor and donor dipole moments respectively. $\Theta = \int D_i(\nu)A_j(\nu)d\nu$ is the overlap integral between single donor fluorescence and acceptor absorbance spectra with intensity normalized to unity, and is expected to be dependent on temperature. The estimated resonant transfer rate in CdSe quantum dots is (38 ps)-1 [12]. The Förster radius, $R_0$ is given by [10]:

$$R_0^6 \propto \frac{Y_D}{n^4}\int D(\nu)\varepsilon_A(\nu)\frac{d\nu}{\nu^4}\ldots \quad (1)$$

where $Y_D$ is the temperature-dependent donor quantum efficiency, $n$ is the refractive index of the film, and $\varepsilon_A$ is the molar extinction coefficient for acceptor absorption. The Förster radius is related to the energy transfer time $\tau_F$ through $\tau_F = \tau_D(R/R_0)^6$ [18] where $\tau_D(T)$ is the lifetime of the donor. For CdSe and PbS quantum dots at lower wavelengths, the Förster radii at room temperature have been estimated at 4.7 [14] and around 8 [18] nm respectively. From our observations, we infer a Förster radius of 12-13 nm at room temperature, based on an interdot separation distance of 10 nm. We note that the discrete "wideband" 12-nm filters used, in comparison with a continuously-tuned narrowband filter, already capture both the Förster photoluminescence quenching and enhancement process, and provide more signal counts for the lifetime measurements.

The FRET efficiency is defined as $\varepsilon^{-1} = 1 - \tau_{DA}/\tau_D$, where $\tau_{DA}$ and $\tau_D$ are donor lifetimes with and without a neighboring acceptor dot. The FRET efficiency at room temperature for donor quantum dots at 1400 nm is estimated at 82% (50% for 1450 nm dots).

We observe comparable efficiencies for quantum dot solids derived from a mixture of two monodisperse QD solutions. Mixed assemblies may allow us to enhance Förster energy transfer rates through better availability of acceptors and donors, as well as through a higher overlap of their corresponding wavefunctions. In Figure 4a and b, we show the results of experiments on a sample derived from a mixture of dots emitting at 1380 nm and 1480 nm, with the wavelengths chosen for enhanced overlap between donor emission and acceptor absorption spectra at the peaks. In the mixed solid, we observe signatures of the two QD samples at room temperature, although there is stronger emission from the larger dots in the solid compared to dots in the mixed solution. An initial increase in population of excitons immediately after excitation is observed at the peak emission of the larger dots. Emission at the peak of the smaller dots also shows distinct donor-type behavior. The efficiency for donor dots, averaged over the entire inhomogeneously broadened spectra is measured to be around 50%.

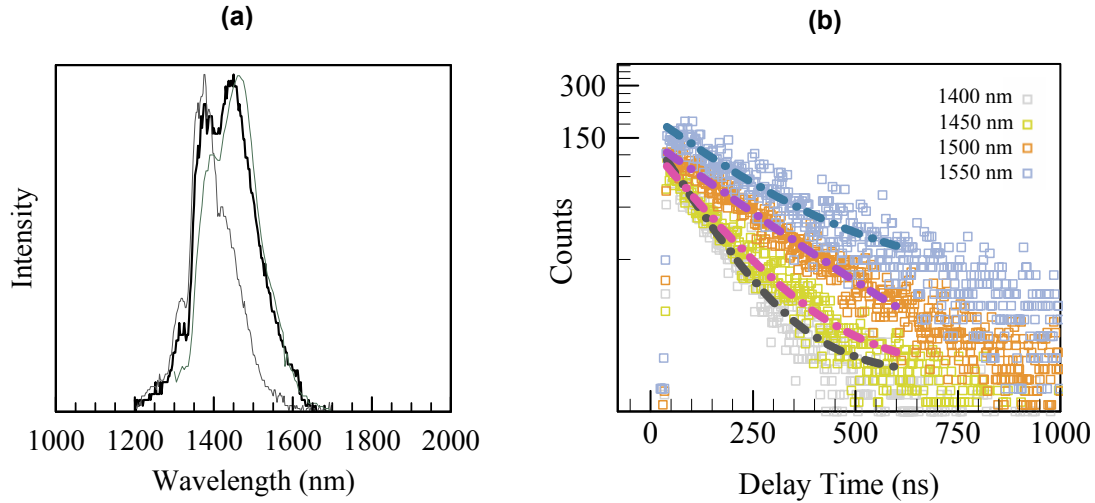

Figure 4: Förster energy transfer between mixed samples: (a) photoluminescence spectra of samples with peaks at 1380 nm and 1475 nm, as well as a QD solid assembled from a mixed solution of the two. (b) Lifetime spectra measured at 1400-, 1450-, 1500- and 1550 nm for the mixed solid, showing distinct evidence of resonance energy transfer. The measured transfer efficiency is ~50%. Dashed curves are monoexponential fits to the measured data ($y = Ae^{-x/\tau} + y_0$).

We find that the efficiency of energy transfer can be further improved in monodisperse samples by temperature-tuning the quantum-dot solid. In experiments, we first map out the photoluminescence lifetime of ensemble quantum dots over the temperature range of 160 K to 295 K (below 160 K, low signal counts from the quantum dots does not permit wavelength-dependent measurements of the photoluminescence lifetime). We observe a lengthening of the photoluminescence lifetime with a decrease in temperature. We also examine time-resolved photoluminescence with 12-nm band pass filters at 160-, 200-, and 295 K. The spectra are shown in Figure 5a.

At 200 K, the Forster energy transfer becomes more pronounced at 1550 and 1600 nm. At 160 K, the photoluminescence decay at 1600 nm shows a remarkably large initial population of dots immediately after excitation. Correspondingly, a very fast quenching of emission for the smallest dots is seen immediately after excitation.

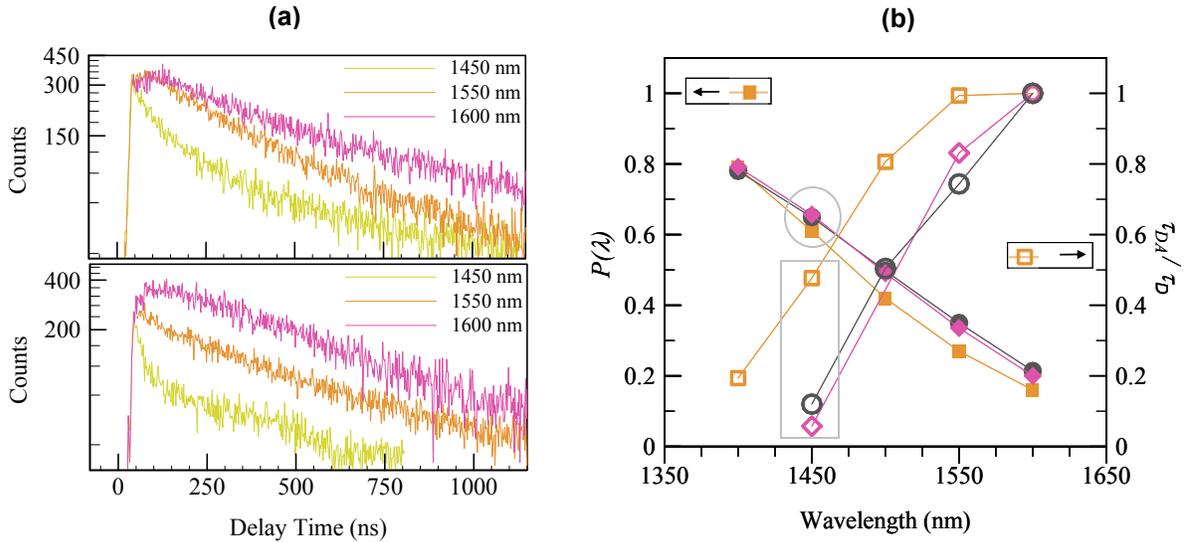

Figure 5: (a) Time-resolved lifetime measurements of PbS QDs in the solid at 200 K (top) and 160 K (bottom). Spectrally-filtered lifetimes at 1450 nm, 1550 nm and 1600 nm demonstrate strong evidence for distinctive Förster energy transfer from smaller to larger dots. The measured Förster energy transfer rate for 1450 nm dots is $(50 \text{ ns})^{-1}$ at 200 K and $(30 \text{ ns})^{-1}$ at 160 K (derived from the initial decay of the small dots). (b) Comparison of Förster energy transfer rates for varying temperature. Filled shapes represent $P(\lambda)$ ($\Delta\lambda_{FRET} = 0$) at 295- (orange squares), 200- (gray circles), and 160 K (pink diamonds), derived as an approximation from the PL spectra of the solid in Figure 3a. Open shapes represent $\tau_{DA}/\tau_D$ where $\tau_{DA}$ and $\tau_D$ represent donor lifetimes with and without the presence of an acceptor respectively. The comparison is highlighted at 1450 nm, where the transfer efficiency increases from 52% to 94% as the temperature is tuned between 295 K and 160 K.

The lifetimes of donor dots at 200- and 160 K are measured to be 50 ns and 30 ns respectively at 1450 nm, leading to estimated FRET efficiencies of 85% and 94% for donor dots at this wavelength. The change in FRET efficiency with temperature in the monodisperse assembly is indeed remarkable. In Figure 5b, we plot the changes in the FRET efficiency as a function of wavelength with changes in temperature. We plot $P(\lambda)$ on the same axes to make reasonable comparisons at different temperature. We especially note that although $P(1450 \text{ nm})$ changes only slightly with temperature, the efficiency increases considerably. This result suggests the possibility of a smaller $\Delta\lambda_{FRET}$ required for resonant energy transfer at reduced temperatures, thereby resulting in a higher availability of acceptor dots. For dots emitting at 1600 nm, the lifetime spectra at 160 K similarly suggest a higher availability of donor dots compared to room temperature.

In conclusion, we report experimental evidence of highly efficient Förster energy transfer in densely packed monodisperse PbS quantum dots in a solid film emitting at 1.5 μm, over a range of 160 K to 295 K. For the smaller dots, we observe a fast initial lifetime component and an eventual relaxation approaching the ensemble lifetime, due to Förster energy transfer. For the larger dots, an initial increase in emission is observed, followed by monoexponential recombination. The resonance energy transfer rate increases from $(128 \text{ ns})^{-1}$ at room temperature to $(30 \text{ ns})^{-1}$ at 160 K for dots emitting at 1450 nm, and is attributed primarily to an improvement in the average functional overlap between donor and acceptor emission and absorption spectra, possibly due to a reduction in the required $\Delta\lambda$ between them. The measured FRET efficiency of 94% at 160 K shows the promise of monodisperse PbS solids for near infrared solar photovoltaics applications, and for examining the control of spontaneous emission lifetimes in QD thin-films through a modified local density of optical states.

We acknowledge helpful discussions with S. Hoogland, E. Sargent, C. Lu, B. Nyman, assistance from S. Jockusch and N. Turro, and funding support from DARPA YFA, the New York State Office of Science, Technology and Academic Research, and the National Science Foundation.